\documentclass[useAMS,usenatbib,aas_macros]{mn2e}

\usepackage{epsfig}
\usepackage{lscape,graphicx,colortbl}
\usepackage{amssymb}
\usepackage{natbib}
\usepackage{times}
\usepackage{aas_macros}

\newcommand{\ifm}[1]{\relax\ifmmode#1\else$\mathsurround=0pt#1$\fi}
\newcommand{\kms}{\ifmmode\,{\rm km}\,{\rm s}^{-1}\else km$\,$s$^{-1}$\fi}

\newcommand{\hmsun}{\,\ifm{h^{-1}}{\rm M_{\odot}}}
\newcommand{\msun}{\rm M_{\odot}}

\newcommand{\dd}{{\rm d}}

\newcommand{\be}{\begin{equation}}
\newcommand{\ee}{\end{equation}}
\newcommand{\bea}{\begin{eqnarray}}
\newcommand{\eea}{\end{eqnarray}}
\newcommand{\z}{\emph{z}}

\def\msb{m_{\rm s, burst}}
\def\mc{m_{\rm cold}}

\def\mb{m_{\rm BH}}

\def\Lb{L_{\rm AGN}}
\def\fp{\alpha_p}
\def\fb{\alpha_{acc}}
\def\lE{\lambda_{\rm Edd,0}} 
\def\sE{\sigma_{\rm Edd}}

\begin{document}

\title[What triggers black-hole growth?]
      {What triggers black-hole growth? Insights from star formation rates}

\author[E. Neistein and H. Netzer]
{Eyal Neistein,$^{1}$\thanks{E-mail:$\;$eyal@mpe.mpg.de} Hagai Netzer$^{2}$
\\ \\
$^{1}$ Max-Planck-Institute for Extraterrestrial Physics, Giessenbachstrasse 1, 85748 Garching, Germany\\
$^{2}$ School of Physics and Astronomy, The Sackler Faculty of Exact Sciences, Tel-Aviv University, Tel-Aviv 69978, Israel\\
\\
}


\date{}
\pagerange{\pageref{firstpage}--\pageref{lastpage}} \pubyear{2013}
\maketitle

\label{firstpage}


\begin{abstract}
We present a new semi-analytic model for the common growth of black holes (BHs) and galaxies within a 
hierarchical Universe. The model is tuned to match the mass function of BHs at $\z=0$ and the luminosity 
functions of active galactic nuclei (AGNs) at $\z<4$. We use a new observational constraint, which relates the
luminosity of AGNs to the star-formation rate (SFR) of their host galaxies.
We show that this new constraint is important in various aspects:
a) it indicates that BH accretion events are episodic;
b) it favours a scenario in which BH accretion is triggered by merger events of all mass ratios; 
c) it constrains the duration of both merger-induced star-bursts and BH accretion events.
The model reproduces the observations once we assume that only 4 per cent of the merger events trigger
BH accretion; BHs accretion is not related to secular evolution; and only a few per cent of the mass made
in bursts goes into the BH. We find that AGNs with low or intermediate luminosity are mostly being triggered by 
minor merger events, in broad agreement with observations. 
Our model matches various observed properties of galaxies, such as the stellar mass function 
at $\z<4$ and the clustering of galaxies at redshift zero. This allows us to use galaxies as a reliable 
backbone for BH growth, with reasonable estimates for the frequency of merger events. Other modes of BH
accretion, such as disk-instability events, were not considered here, and should be further examined in the
future.
\end{abstract}




\section{Introduction}
\label{sec:intro}

Most massive galaxies in the local Universe host supermassive black holes (BHs) in their 
centres \citep{Kormendy95}. Active periods of accretion onto such objects, seen as active
galactic nuclei (AGNs), are observed in less than a few percent of the objects. 
Consequently, models of BH evolution rely on assumptions about the triggering of accretion 
events and the connection, if any, with the host galaxy evolution.  A key problem in such 
studies relates to the big difference in scale between the cold gas in the galaxy 
(distributed over several kpc) and the inner disk around the BH (typically smaller than a parsec). Various 
suggestions have been made to bridge the gap between these scales, including: 
merger triggered accretion, secular processes in the host galaxy including disk 
instabilities\footnote{In this work we use the terms ``secular evolution'' and ``secular processes'' to indicate
all types of internal processes within a galaxy. These include events of disk
instability that might induce bursts of star-formation. Secular evolution does not include events 
related or triggered by galaxy mergers.}, 
mass loss from evolved stars, star formation (SF) activity in the central region, and more.

The merger scenario is motivated by two different lines of evidence. First, $N$-body and hydrodynamical simulations 
of merging galaxies suggest that such events drive a large amount of cold gas into the centre of the merging 
system, resulting in the formation of a large bulge or a spheroidal galaxy \citep{Barnes91, Mihos96, Cox06, 
Robertson06, DiMatteo07}. Second, AGN activity is observed in local ultra-luminous infrared galaxies (ULIRGs),
which are known to be associated with energetic merger events \citep{Sanders88, Sanders96, Surace00, Canalizo01, 
Veilleux09}. Observationally, an evidence for a direct connection between minor mergers and AGN activity
is more difficult to establish.

Recent studies show that most galaxies at $\z<2$ that host a low-luminosity AGN
do not show morphological evidence for major 
merger events in their recent history \citep{Gabor09, Cisternas11, 
Kocevski12, Schawinski12}. Only high-luminosity AGNs are 
related to major mergers \citep{Treister12}. This has been taken to indicate that the role of secular 
processes in driving gas to the centre of galaxies
and feeding the BH is important and very common at almost all redshifts 
\citep{Efstathiou82, Genzel06, Elmegreen10, Ceverino10, Bournaud12}. 
The nature and frequency of such internal events are not fully understood, partly
because it is difficult to discriminate them from minor mergers. 
Nonetheless, various structural properties of galaxies have been suggested to distinguish
between secular and merger events \citep{Kormendy09}.

Most of the studies mentioned above have measured
the mass ratio in mergers by visual inspection, involving a large uncertainty.
Moreover, there may well be a time-delay between the merger event and the peak luminosity of the AGN. 
Since recently merged galaxies have similar clustering properties as other galaxies of the same mass 
\citep{Bonoli10}, the effect of time-delay is difficult to detect \citep[however, it might
be constrained by detailed analysis of galaxy spectra][]{Davies07, Wild07, Wild10}.

Other internal mechanisms within a galaxy were proposed as a trigger for BH accretion. \citet{Hopkins06a} 
have demonstrated that feedback from the BH can regulate low levels of BH accretion. \citet{Best05} have argued 
that observations of radio-loud AGNs provide indications for hot gas accretion into BHs.

The mass of super-massive BHs in galactic centres is strongly correlated with both the bulge mass 
\citep{Magorrian98, McLure02, Marconi03, Haering04, Sani11, McConnell12, Graham12, Graham13} and the bulge velocity dispersion 
\citep{Ferrarese00, Gebhardt00}. These relatively tight correlations suggest that the mechanisms which are responsible
for building up the mass in bulges are also related to BH growth and hence the triggering of AGN activity. 
However, since the mass is an integrated
quantity, these correlations could also originate from the hierarchical aggregation of mass 
\citep{Hirschmann10,Jahnke11}. Better understanding of the various accretion scenarios is therefore 
required in order to understand the BH-bulge relationships.

Further hints for the nature of BH accretion could be provided by relating AGN luminosities to SF  
rates (SFRs) of their host galaxies. Unlike morphological studies, SFRs are well defined, and can
be quantitatively compared between various populations of galaxies. Studies of SFRs and AGNs 
luminosities in large spectroscopic surveys, like the Sloan Digital Sky Survey \citep[SDSS,][]
{York00} suggest that many AGNs are located within star-forming galaxies 
\citep{Kauffmann03a, Wild07, Salim07, Daddi07, Davies07, Silverman09, 
Netzer09, Wild10, Mor12, Rovilos12, Tommasin12}. Here we choose to highlight the recent measurements by 
\citet{Shao10} and \citet{Rosario12} that are based on SFRs obtained by {\it Herschel} in the FIR, and are more 
reliable than UV-based SFRs. According to \citet{Rosario12}, such SFRs are uncorrelated with the AGN luminosity 
at low and intermediate luminosities and at high redshift. However, 
\emph{high} luminosity AGNs at low and intermediate redshifts do show a correlation with the SFRs of their host 
galaxies.

In this paper we use a semi-analytic model (SAM) of galaxy-formation combined with BH evolution, in order
to examine the correlations between SFRs and AGN luminosities. SAMs are naturally
being used for this problem as they provide a statistical sample of galaxies that is in general
agreement with observations \citep{Kauffmann00, Croton06, Bower06, Malbon07, Monaco07, Somerville08, 
Bonoli09, Fanidakis11, Fanidakis12, Hirschmann12}.

Our study is different from previous SAMs in various
aspects. First, we use the SAM from \citet{Neistein10} and \citet{Wang12} that fits the stellar mass function 
of galaxies at $0<\z<4$ and the clustering of galaxies at low redshift. As a result, the star-formation 
histories and merger-rates of galaxies within the model are in broad agreement with the observed Universe.
Second, we do not attempt to model AGN feedback as a function of each specific AGN but rather assume
that the galaxies within our model experience an average AGN feedback that depends only on halo mass.
Third, we aim to match a wide range of AGN observations, including the relation 
between SFR and AGN luminosity mentioned above, and the fraction of host galaxies that are experiencing 
major merger events.

Various recent empirical models of BHs and AGNs evolution are aimed to fit the BH mass 
function and AGN luminosity functions \citep[e.g.][]{Wyithe03, Lapi06, Hopkins06b, 
Hopkins07, Hopkins08, Croton09, Shankar09, Shankar10, Shankar12, Pereira11, Conroy12, Draper12}. 
Here we choose to use SAMs, as they allow a more direct modelling of SFRs
and merger fractions of individual objects, in close contact with the formation history of their host haloes.
These features play a key role in our model.

This paper is organized as follows. In section \ref{sec:model} we describe the model ingredients and 
the details of computing all the properties of galaxies and BHs. 
Section \ref{sec:results} includes a detailed comparison between the model and observations.
The results are summarized and discussed in section \ref{sec:discuss}.
This study is based on the cosmological parameters that are used by the Millennium simulation:
$(\Omega_m,\Omega_\Lambda,\sigma_8,h)=(0.25,0.75,0.9,0.73)$. Throughout the paper we use $\log$ to
denote $\log_{10}$.


\section{The model}
\label{sec:model}

\subsection{Galaxy formation and evolution}
\label{sec:model_gal}

The SAM used in this work is adopted from \citet{Wang12}, and is based
on the formalism of \citet{Neistein10}. The model follows galaxies inside the complex structure
of subhalo merger-trees taken from a large $N$-body cosmological simulation \citep[the Millennium
simulation,][]{Springel05}. This simulation follows 2160$^3$ dark-matter particles inside a periodic box of 
length 500 $h^{-1}$Mpc, with a minimum halo mass of $1.72\times 10^{10} \hmsun$.

Our SAM includes the effects of cooling, star formation (SF), accretion, merging, and feedback. 
Unlike other SAMs, these laws are simplified to be functions of only the host subhalo mass and redshift. 
We have shown in \citet{Neistein12} that this concept is sufficient for the SAM to reproduce the
gas and stellar mass content of galaxies, on an object by object basis, as obtained
from a hydrodynamical simulation to an accuracy of 0.1 dex. Consequently, this SAM is 
complex enough to accurately follow the SF histories of galaxies.

The specific model used here was presented in \citet{Wang12} as model 4. It is based on
the SAM of \citet{DeLucia07}, transformed to the language of \citet{Neistein10}
and including the following modifications:
\begin{itemize}
\item SF efficiency (the ratio between the SFR and the cold gas mass within the disk) 
is lower in small mass haloes.
\item Satellite galaxies experience stripping of hot gas in proportion to the stripping of 
dark-matter, following the suggestion of \citet{Weinmann10} \citep[see also][]{Khochfar08}.
\item Cooling is suppressed within haloes that are more massive than $5.6\times10^{11}\hmsun$, and
there is no SF within haloes of mass $>5\times10^{12}\hmsun$ at $\z<1.3$. These ingredients are aimed to
mimic the feedback from AGNs, although they do not depend on the specific AGN hosted by each galaxy.
\item The dynamical friction time is assumed to depend on the cosmic time and is shorter at higher redshift.
This behaviour is obtained by using the Chandrasekhar formula with a multiplication factor of
$5\times(t/13.6)^{0.5}$, where $t$ is the time in Gyr since the big-bang. This dependence 
is motivated by a more radial
infall of satellite galaxies towards the centre of their group at high redshifts \citep{Hopkins10,Weinmann11}.
\item The SFR in merger induced bursts depends on the halo mass. In particular, it is lower than the approximation 
used by \citet{DeLucia07} for haloes less massive than $3\times10^{11}\hmsun$. The specific implementation is 
described below.
\end{itemize}

Galaxies in our model are embedded within dark-matter subhaloes as extracted from the $N$-body simulation. 
We assume that the galaxies merge following the merging-time of their host subhaloes, 
with an additional time-delay.
This delay is estimated using the Chandrasekhar formula for dynamical friction \citep[see][]{Wang12}. 
Mergers trigger SF bursts, with an efficiency that depends on the mass ratio of the two galaxies. The total
stellar mass formed in the merger induced burst is:
\begin{equation}
\label{eq:merger_effic}
\Delta \msb = \left\{ \begin{array}{ll}
0.56 \left(\frac{m_2}{m_1}\right)^{0.7} \mc & \;\;\;\textrm{if } M_h\geq M_0 \\ \;\;\;\
& \;\;\;\;\;\;\;\;\;\;\;\;\;\;\;\;\;\;\;\;\;\;\;  \\
0.56 \left(\frac{m_2}{m_1}\right)^{0.7}  \mc  \frac{M_h}{M_0} & \;\;\; \textrm{otherwise}
\end{array} \right.
\end{equation}
where $m_1\,,m_2$ are the baryonic masses of the central and satellite galaxy respectively (including
both stellar and cold gas mass), $\mc$ is the sum of the cold gas masses of the two galaxies, $M_h$ is 
the mass of the descendant subhalo, and $M_0=3\times10^{11}\hmsun$. 
For high mass galaxies, this recipe follows the results of hydrodynamical simulations by 
\citet{Mihos94} and \citet{Cox08}, and was adopted by various SAMs \citep[e.g.][]{Somerville01, Croton06, 
Khochfar09, Neistein10, Khochfar11}. For low mass galaxies this recipe is motivated by recent observations
of SF efficiency \citep{Wang12}. Note that mergers provide the only trigger for 
SF bursts in our model. The other mode of SF, by secular processes,  
describes the relatively slow conversion of cold gas into stars within disk galaxies.

The specific model used here is able to fit the SF histories of 
galaxies to a higher accuracy than most other SAMs \citep[see][for
a recent success in this respect]{Henriques12}. As was shown in \citet{Neistein10},
our models reproduce both the stellar mass function of galaxies up to $\z=4$, and the distribution of 
SFRs at $\z=0$. In \citet{Wang12} we further improved the model to reproduce the auto-correlation function
of galaxies at $\z=0$. These results are unique in comparison to other SAMs, allowing us
to use reliable SF histories of galaxies as a basis for the
work presented here which focuses on BH evolution.
The model fits observations to a level of 20-40 per cent, and is probably not
the only possible galaxy formation model. In particular, there is a degeneracy between the amount of 
stars formed in mergers induced bursts, versus the amount of star formation due to secular processes.

The duration of SF bursts has a negligible effect
on the statistical properties of the model galaxies. This is because these events are rare, 
and hardly contribute to
the total SFR density of the Universe. However, in our current study, bursts are the only
channel of BH growth, and their duration plays an important role in the model. We 
assume that the shape of SF bursts follows a Gaussian as a function of time, with the following parameterization:
\begin{equation}
\label{eq:merger_time}
\frac{\dd\msb}{\dd t} = \frac{\Delta\msb}{\sigma_b \sqrt{2\pi}} \exp\left[-\frac{(t-t_0)^2}{2\sigma_b^2}\right] \,.
\end{equation} 
Here $\sigma_b$ determines the burst duration, $t$ is the time in Gyr since the big-bang, and $t_0$ is the time of the
peak in the SF burst. We assume that $t_0$ occurs $2\sigma_b$ after the time the galaxies merge, 
to allow a smoothly rising peak of SFR (the burst of SF starts at the time of merging). 
We allow the burst to continue forming stars up to $4\sigma_b$ after $t_0$, 
so the total duration of the burst is $6\sigma_b$.

At higher redshifts galaxies have smaller radii, and 
time-scales are shorter (the halo dynamical time is proportional to the cosmic time, $t$).  We therefore assume 
that $\sigma_b$ depends on $t$ in the following way:
\begin{equation}
\label{eq:sigma_b}
\sigma_b = \sigma_0  \frac{t}{13.6} \,\, {\rm Gyr}\,,
\end{equation}
where $\sigma_0$ is a free constant.

An additional important ingredient of the model is the definition of the bulge mass.
We assume that galaxies can grow a bulge according to two different channels. 
First, SF bursts that are triggered by mergers of any mass ratio 
are assumed to contribute their stellar content to the bulge (i.e., $\Delta\msb$ from 
Eq.~\ref{eq:merger_effic} is added to the bulge of the remnant galaxy). 
Second, once the mass ratio $m_2/m_1$ is larger than 0.3,
the \emph{total} stellar mass of both galaxies is moved to the bulge of the remnant galaxy.
Therefore, the mass of stars within the bulge might be larger than the total amount of stars formed within bursts.

\subsection{Black holes}
\label{sec:model_bh}

Our SAM assumes that BHs grow only during merger induced star bursts. 
The masses of seed BHs in this model are extremely small in comparison to the mass added in the first
merger event and make no difference to the accumulated mass of BHs at relatively low redshifts. 
We have tested that using BH seeds of mass $10^3\,\msun$ does not change our model results significantly. 

When galaxies merge, we merge their corresponding BHs at the same time. Following each merger event, a burst of SF 
occurs according to Eqs.~\ref{eq:merger_effic}-\ref{eq:sigma_b}. We allow the remnant BH to grow in mass according to:
\begin{equation}
\label{eq:bh_growth}
\frac{\dd\mb}{\dd t} = \left\{ \begin{array}{ll}
(1-\eta) \frac{\dd\msb}{\dd t}  \fb & \;\;\;\textrm{with probability } \fp \\ \;\;\;\
& \;\;\;\;\;\;\;\;\;\;\;\;\;\;\;\;\;\;\;\;\;\;\;  \\
0 & \;\;\; \textrm{otherwise}
\end{array} \right.
\end{equation}
Here $\eta$ is the fraction of mass that is transformed into radiation and $\fb$ is a free parameter, 
corresponding to the efficiency of BH accretion with respect to the SF burst. 

In a $\Lambda$CDM universe, the fraction of galaxies with an active SF burst at any given time is high, 
especially due to the non-negligible burst duration $\sigma_b$. However, observations show that the number of 
AGNs is significantly lower \citep[e.g.][]{Croom09}. Our model 
assumes that only a fraction $\fp$ of the merger events induce accretion into the BH. In practice, at each time-step 
of the SAM (with a typical duration of 10 Myr) and for each BH we generate a random number, distributed uniformly 
between zero and unity. We then allow accretion only if this random number is smaller than $\fp$.
In practice, each merger event adds on average a mass of $\fb \fp\Delta\msb$ to the BH.

The bolometric luminosity of the AGN is defined as:
\begin{equation}
\Lb = \frac{\eta}{1-\eta} \frac{\Delta\mb}{\Delta t} c^2 \,,
\end{equation}
where $\Delta t$ is the length of the time-step within the SAM ($\sim10$ Myr), $\Delta\mb$ is the total mass that is
added to the BH within a time-step (i.e., the integration of Eq.~\ref{eq:bh_growth}), and $c$ is the speed of light. 

The Eddington luminosity describes the limit at which radiation pressure balances the gravitational force
of the BH
\begin{equation}
L_{\rm Edd} = 1.5 \left( \frac{\mb}{10^8 \msun}\right) \, 10^{46} \,\,\, {\rm \,erg \,s^{-1}} \,,
\label{eq:Ledd}
\end{equation}
where the factor 1.5 on the left is derived for solar metallicity gas.
In our model, the accretion into the BH is determined by the SF burst, and might be above the 
Eddington limit. In addition, it is reasonable to assume that this limit will vary between different BHs.
In order to use a practical accretion limit,
we define $\lambda_{\rm Edd}$ to be a log-normal random variable (i.e. its log value is normally distributed) 
with a mean of $\lambda_{\rm Edd,0}$ and a standard deviation of $\sigma_{\rm Edd}$ (in log). 
For each accretion event, we randomly choose a value for 
$\lambda_{\rm Edd}$ and do not allow $\Lb$ to exceed $L_{\rm Edd} \lambda_{\rm Edd}$. We will refer
to $L_{\rm Edd} \lambda_{\rm Edd}$ as the Eddington \emph{threshold}, since it is being used to actively
limit the accretion onto BHs. We note that we do 
not use the Eddington threshold for the first accretion event (i.e. when the BH mass equals zero).

At low accretion rates we use the properties of `advection dominated accretion flows' (ADAFs). 
As in previous studies \citep[e.g.][]{Fanidakis12} we model the ADAF limit by
\begin{equation}
\eta_{\rm ADAF} = \frac{\Lb}{L_{\rm Edd}} \frac{\eta}{\alpha_{\rm ADAF}}
\label{eq:adaf}
\end{equation}
which is valid only for $\Lb/L_{\rm Edd}<\alpha_{\rm ADAF}$. By definition, $\eta_{\rm ADAF}$ equals $\eta$
when $\Lb/L_{\rm Edd}=\alpha_{\rm ADAF}$. In this work we fix $\alpha_{\rm ADAF}$ to a value of 0.01.

Some of the results presented here make use of the luminosity of AGNs in the $B$ band. For the conversion between
$\Lb$ and $B$ band we use the bolometric correction from \citet{Marconi04}:
\begin{equation}
\log \left( \nu_B L_{\nu_B}\,/\, \Lb \right) = -0.8 + 0.067L_1 - 0.017 L_1^2 +0.0023L_1^3 \,.
\end{equation}
where $L_1 =\log(\Lb/L_\odot)-12$. The absolute magnitude in the AB system is then given by
\begin{equation}
M_{B} = -11.32 - 2.5 \log\left( \nu_B L_{\nu_B}/10^{40} {\rm \,erg \,s^{-1}} \right) \,,
\end{equation}
where $B$ refers to a rest wavelength of 4400\AA.

To summarize, our SAM includes all the ingredients from \citet{Wang12} related to galaxies, with the following
additional parameters related to the evolution of BHs: $\eta$, $\fb$, $\fp$, $\lambda_{\rm Edd,0}$, and
$\sigma_{\rm Edd}$. 
These parameters are presented in Table \ref{tab:parameters} and their best chosen values are justified in 
section \ref{sec:results}.
The ingredients used here are similar to previous SAMs \citep[e.g.][]{Malbon07}, although we do not
include growth modes that are not due to bursts \citep[e.g.][]{Hirschmann12}.

Finally, we note that the model is
limited both in terms of the minimum BH mass that is properly resolved,
and in terms of the limited volume of our simulation box. These limitations originate from the limited dynamical
range of the Millennium simulation that is being used here, and will be discussed below.


\section{Results}
\label{sec:results}

\subsection{Observational constraints and best model parameters}

\begin{table*}
\caption{The free parameters used to model BHs and AGNs. }
\begin{center}
\begin{tabular}{llll}
\hline Parameter & range & best value & description  \\
\hline
$\sigma_0$  &  0.1-0.5 Gyr  & 0.13 Gyr & Burst duration, Eq.~\ref{eq:sigma_b} \\
$\eta$  &  0.05-0.3 & 0.05 & Radiative efficiency, Eq.~\ref{eq:bh_growth}  \\
$\fb$  &  0.001-0.3 & $0.185(t_1^2 - 1.68t_1 + 0.75)$ & Accretion efficiency, Eq.~\ref{eq:bh_growth}, 
$t_1=t/13.6$ Gyr  \\
$\fp$ & 0.001-0.1 & 0.04 & Fraction of mergers that result in AGN activity, Eq.~\ref{eq:bh_growth}  \\
$\lE$ & 0.1-3 & 1.0 & Eddington threshold (mean)     \\
$\sE$ & 0-1 & 0.7 & Eddington threshold (scatter)   \\
\hline
\end{tabular}
\end{center}
\label{tab:parameters}
\end{table*}

In this work we treat the galaxies as priors
and do not change the parameters that affect their evolution, except for the value
of $\sigma_0$, which has a negligible effect on the properties of galaxies. Consequently, there are only six free 
parameters in the model, as listed in Table \ref{tab:parameters}. However, the low
density of AGNs in the Universe forces us to test each set of model parameters by using the full Millennium simulation,
spending a few CPU hours on each run.  
In order to achieve fast tuning, while still running our model on the full simulation box, we first save all the 
model results that are related to the evolution of galaxies. We then run the BH ingredients only, thus saving a 
large amount of computational time. Our tuning procedure can evolve BHs over the full Millennium simulation 
in only three minutes (using one processor), allowing us to systematically explore a large region of the parameter 
space.

\begin{figure}
\centerline{ \hbox{ \epsfig{file=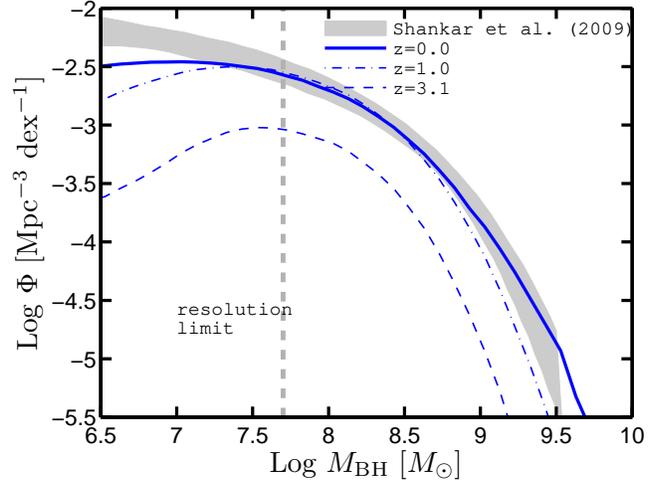,width=9cm} }}
\caption{The mass function of BHs. The \emph{shaded} region represents the observational prediction
from \citet{Shankar09} based on the correlation between bulge and BH masses at low redshift. 
Different \emph{line} types correspond to the model mass function at various redshifts as indicated.
The grey \emph{dashed} line shows the minimum BH mass that is reliably reproduced by the model 
(due to a minimum subhalo mass of $1.72\times10^{10}\,\hmsun$).}
\label{fig:bh_mf}
\end{figure}

\begin{figure*}
\centerline{\psfig{file=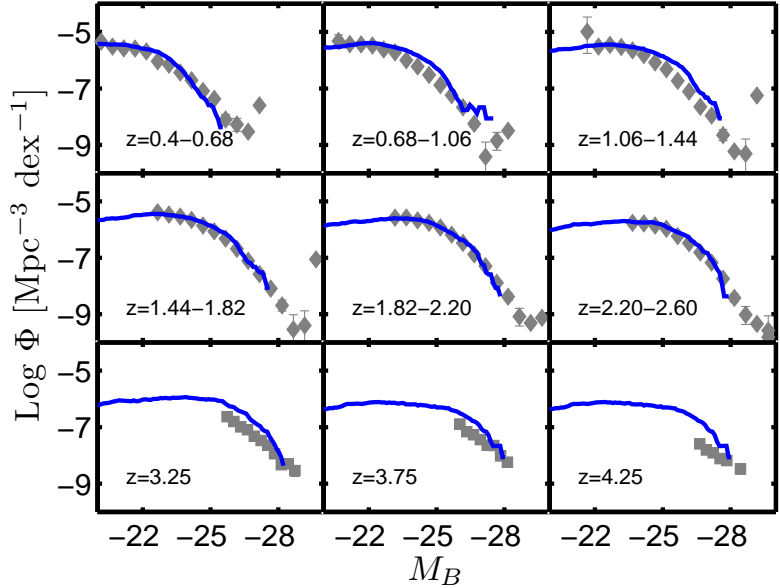,width=120mm,bbllx=20mm,bblly=90mm,bburx=168mm,bbury=200mm,clip=}}
\caption{The luminosity functions of AGNs. Each panel shows one redshift bin as indicated.
Observational results from \citet{Croom09} and \citet{Richards06} are plotted in 
\emph{diamonds} and \emph{squares} symbols respectively. 
Following Eq.~2 and 3 from \citet{Croom09}, we add 0.587 and 0.817 magnitudes to $M_g$ and $M_i$ 
respectively, in order to convert them to $M_B$ 
(note that in these papers $M_g$ and $M_i$ are defined for rest wavelengths corresponding 
to $\z=2$ and hence require correction factors to compare with our computed $B$ band at a wavelength of
4400\AA). Model results are plotted in \emph{solid} lines.}
\label{fig:bh_lf}
\end{figure*}

We use three different types of observations to constrain our model parameters: the luminosity functions
of AGNs at $0<\z<4$, the mass function of BHs at $\z=0$, and the relation between SFRs and AGN luminosities.
While we further compare the model to other observations in the following sections, these additional 
observations were not used for tuning the model.

\subsubsection{The mass function of BHs and the luminosity functions of AGNs}

We start by tuning the model parameters against two common observations: 
the mass function of BHs at redshift zero, and the luminosity function of AGNs at $\z<4$. 
We apply a tuning procedure that systematically scans a range of the parameters $\fb$, $\fp$, $\lE$, $\sE$
according to the range listed in Table \ref{tab:parameters}.
The value of $\sigma_0$ can significantly affect the luminosity function of AGNs. However, it is fixed here 
due to its effect on the relation between SFRs and AGN luminosities, that will be discussed in the next section. 
Using the prior value of $\sigma_0=0.13$ Gyr, we could not find a model 
that matches the data to a satisfying accuracy, while using constant values for all the parameters. 
We note that larger values of $\sigma_0$ allow for better models. Consequently, we choose to add a dependence
on time for $\fb$. It should be noted that a similar result as shown here could arise from varying 
the parameter $\fp$ with time.
We will discuss below the degeneracy in choosing the best solution, and its relation
with the SFR values of our model galaxies. Our best model uses $\eta=0.05$,
consistent with a value of 0.057 obtained for non-rotating BHs.

In Fig.~\ref{fig:bh_mf} we show that the mass function of BHs from the model at $\z=0$ agrees with the prediction from 
\citet{Shankar09}. The results from \citet{Shankar09} are based on the mass function of galaxies, combined
with the correlation between the bulge and the BH mass. It is therefore not a direct observation, and should
not be considered as a strong constraint. However, the mass function points to important features of the model
and we choose to use it as a constraint even though the actual values are not certain. 
For example, the recent work by \citet{McConnell12} indicates that the ratio between the mass of bulges
and BHs could reach a level of $\sim100$ at the high-mass end. Fig.~\ref{fig:bh_mf} also shows 
the model mass functions at higher redshifts, indicating that the number of low mass BHs grows fast 
before $\z\sim1$ and changes only slightly at $\z<1$.

As seen in Fig.~\ref{fig:bh_mf}, the computed mass function declines at 
$\log \mb/\msun\lesssim 7.7$. We have tested a lower resolution model by allowing BHs to grow only
within subhaloes more massive than $10^{11}\hmsun$. This mimics a low-resolution simulation with respect
to the actual minimum mass of the simulation ($1.72\times10^{10}\hmsun$). Within this resolution test, the bend in the
mass function occurs at a larger mass, indicating that our resolution limit is indeed the reason for the bend 
at $\log \mb/\msun\lesssim 7.7$. Our limited resolution has a negligible effect on other results shown in this work.

Models based on high resolution merger trees could possibly solve the above resolution problem
\citep[e.g.][]{Fanidakis12}. Since it is difficult to increase the mass resolution within cosmological
simulations, this issue could be addressed by using Monte-Carlo algorithms for 
generating merger-trees of haloes \citep[e.g.][]{Somerville99a, Parkinson08, Neistein08a}. However, the merger-trees 
used here are based on subhaloes, and are more accurate than the theoretical merger trees based on haloes. We 
plan to check the use of Monte-Carlo merger trees in a future study.

The luminosity functions (LFs) of AGNs are shown in Fig.~\ref{fig:bh_lf}
where points are observations adopted from \citet{Croom09}, \citet{Richards06} and 
solid lines are the model results. The theoretical LFs computed by the SAM are in broad agreement
with the observations. In case we use constant $\fb$ the LFs at $\z<1$ are higher
than the observed ones. We choose to use only optical LFs here since it was shown by previous studies
that different LFs (e.g. based on bolometric luminosity or X-ray), do not add more constraints on the models 
\citep{Hirschmann12,Fanidakis12}. 

Previous studies using the merger scenario as the main mechanism of feeding BHs,
 like the one presented in \citet{Marulli08}, had difficulties matching the
AGN LF. Our model is different from these earlier attempts. In particular, it is based on improved SF histories
which match the observed stellar mass functions over a large range of redshifts.
Such modifications alter the amount of cold gas available for bursts within mergers, and the
number of major versus minor mergers. Consequently, the AGN LF is different although the mode of
BH accretion is the same (see the Appendix for more details).

It should be noted that our simulation volume is smaller than the observed volume at $\z\gtrsim2$, giving
rise to some of the deviations seen at high luminosities at this redshift range.
Furthermore, we do not take into account various effects that could change the model
at the level of $\sim$30 per cent. For example, it is well known that some AGNs are obscured and
would not be accounted for in the observed LFs. 
We have tested that using the slim-disk approximation for super Eddington accretion rates \citep[e.g.][]
{Fanidakis12} does not change the model results.
Lastly, re-tuning the ingredients related to galaxy-formation might further improve the model results. 
We neglect all these effects in order to keep our model as simple as possible.

Most of the free parameters in our model have a simple effect on the LF. Since the SF bursts of
\emph{galaxies} are not modified, $\fb$ behaves as 
a constant multiplication factor for the luminosity of all AGNs at a given redshift.
Changing  $\fb$ thus induces a lateral shift in the luminosity function. The value of $\fp$ changes the number
of bursts leading to BH activity, and therefore shifts the luminosity 
function along the Y-axis. Both $\fb$ and $\fp$ are degenerated at some level, yielding similar results for 
different combinations with the same value of $\fb\fp$. The value of $\eta$ is
similar in its effect to $\fb$.  
In addition, it changes the relations between mass growth (i.e. the mass function
of BHs) and luminosity. The parameters of the Eddington threshold, $\lE$ and $\sE$, 
change mostly the high-end part of the luminosity functions, and also the mass function of BHs. These two parameters 
do not have a simple effect on the model results. 
When neglecting the Eddington threshold, the luminosity of AGNs only depends on the SFR within bursts, and not on 
the BH mass. In this simplified case, the LF of AGNs has the same shape as 
the number-density of SFRs in the model.

\subsubsection{Star-formation rate and AGN luminosity}

\begin{figure}
\centerline{ \hbox{ \epsfig{file=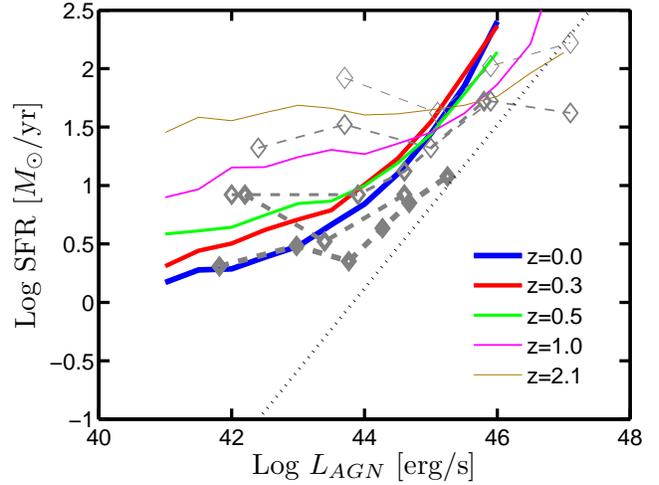,width=9cm} }}
\caption{The average SFR for host galaxies of a given AGN luminosity. Observational results from
\citet{Rosario12} are shown in grey \emph{dashed} lines, where thinner lines correspond to higher redshifts.
The observed error bars are typically at the level of $\pm0.2$ and $\pm0.5$ dex, for SFR
and $\Lb$ respectively.
Model results using the same averaging scheme and the same redshift bins are plotted in \emph{solid} lines 
with redshifts as indicated. For reference, we plot in \emph{dotted} line the relation ${\rm SFR}\propto \Lb^{0.7}$
(the plot is referred to in the paper as the `rates diagram').}
\label{fig:sfr}
\end{figure}

\begin{figure}
\centerline{ \hbox{ \epsfig{file=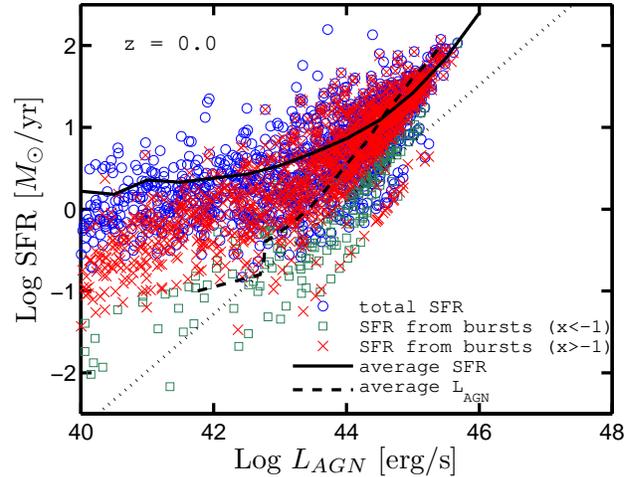,width=9cm} }}
\caption{The rates diagram: values of SFR versus $\Lb$ at $\z=0$. \emph{Circles} represent the SFRs and AGN
luminosities of the model galaxies. \emph{Cross} and \emph{square} symbols show the level of SFR for 
the same objects, when considering contributions from SF bursts only (symbols are plotted for 20 per cent
of the objects). We use the variable $x=(t-t_0)/\sigma_b$, where $x<-1$ corresponds to objects in the 
early stages of the burst. The thick \emph{solid} line shows the average SFR value for a given $\Lb$, as
was computed in Fig.~\ref{fig:sfr} (i.e., based on the total SFRs). 
The \emph{dashed} line corresponds to the average value of $\Lb$, for a given value of total SFR.}
\label{fig:sfr_var}
\end{figure}

In Fig.~\ref{fig:sfr}  we show the observed results from \citet{Rosario12} in grey dashed lines. 
Values along the X-axis correspond to the bolometric luminosity of X-ray selected AGNs. 
The original Y-axis of this plot corresponds to
the rest-frame luminosity at a wavelength of 60 $\mu$m, stacking all galaxies 
that host AGNs with similar $\Lb$. In order to transform the IR luminosity into SFR, we use 
the relation $L_{IR}=3.8 \times 10^{43}$SFR (erg s$^{-1}$) and $L_{60\mu m}=0.5L_{IR}$, 
where SFRs are given in units of $\msun$ yr$^{-1}$.
The corresponding SFR values are thus population averages for various redshift groups and do not represent 
individual objects.

Interestingly, as discussed in \citet{Rosario12},
for $\z\geq1$ there is hardly a correlation between the AGN luminosities and SFRs. 
At $\z<1$ the correlation between SFRs and AGN luminosities only exist for $\Lb>10^{44}$ erg s$^{-1}$.
These findings seem surprising, since we expect that AGN growth modes will be coupled to the properties
of their host galaxies. A lack of correlation might indicate that most AGNs are triggered by processes that
are not related to the dominant mode of SF. In addition, it does not seem obvious that the correlation
for bright AGNs only exist at low redshifts. In what follows, we term this plot as the `rates diagram', since it
shows the relation between BH accretion \emph{rates}, and SF \emph{rates} of their host galaxies.

An explanation for the rates diagram should take into account the
various time-scales that are buried within this relation.
First, the observations of SFRs are based on indicators that last $\sim$150 Myr (this
is a combination of both UV light emission from young stars, and dust heating that yields IR luminosity).
Hence the observed SFR values reflect some of the recent history of the galaxy. To follow the same effect in
the model, we compute the average SFR for each galaxy over time, taking into account the last 150 Myr before the 
output snapshot. Second, AGN luminosities represent an instantaneous 
emission, assumed by the model to last for $\sim10$ Myr.  
Third, the duration of both the burst of SF and AGN activity is $6\sigma_b\approx 0.06t$
(see Eqs.~\ref{eq:merger_time} \& \ref{eq:sigma_b}).

The time-averaged SFR is crucial to our main findings and it was therefore tested under
a different assumption. We have used the stellar bolometric luminosity as a function of 
time as a proxy of the FIR luminosity. The bolometric luminosity is obtained from 
\citet{Bruzual03} for a delta function starburst. This has resulted in only a 
minor change of the results. The reason is that although the stellar bolometric luminosity 
starts to decrease after about $10^7$ years, it remains high enough to affect the integration 
of the SFR over several hundred Myrs.

As a result of the SF time-scale,
the observed value of the IR luminosity is affected 
by the secular SF that occurred before the burst. This luminosity 
is uncorrelated with the merger event and with the AGN luminosity. 
High luminosity AGNs are those that are observed at a time that is close to
the peak of their accretion event ($t_0$ from Eq.~\ref{eq:merger_time}). In these cases the SF burst is significant,
and consists of most of the observed SFR value. This explains why the correlation between SFRs and AGN luminosities
is only seen at high AGN luminosities. At $\z=2$, the secular SFR for star-forming galaxies is high, 
reaching the same level of SFR due to bursts in the most luminous AGNs. This is the reason for the lack of
correlation between SFR and AGN luminosity at these redshifts.

\begin{figure}
\centerline{ \hbox{ \epsfig{file=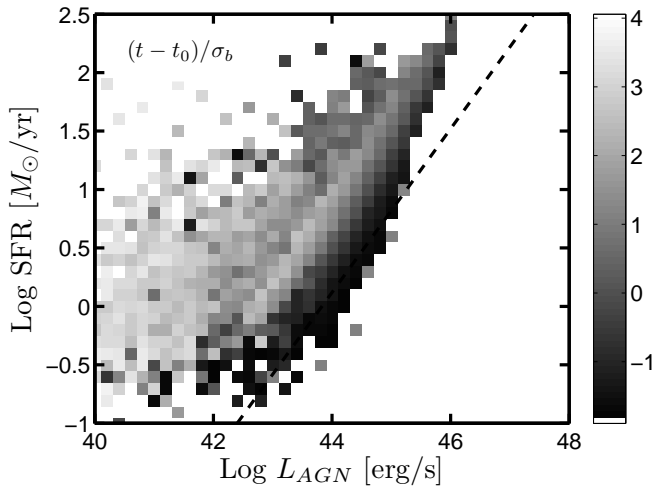,width=9cm} }}
\caption{The temporal location of AGNs with respect to the peak luminosity of the burst.
For each value of SFR and $\Lb$ we plot 
the median value of $x=(t-t_0)/\sigma_b$ (see Eq.~\ref{eq:merger_time}).
All AGNs are selected at redshift zero.
The \emph{dashed} line follows the relation ${\rm SFR}\propto\Lb^{0.7}$. 
AGNs that are starting their accretion episode lie on the right part of the diagram.}
\label{fig:sfr_burst_dt}
\end{figure}

At a given observed epoch $t$, different bursts have different values of $t_0$ and $\Delta \msb$ only, as all other 
parameters do not vary between different objects. Since the time-scale for averaging the AGN luminosity is short,
objects with the same value of $\Delta\msb$ and $|t-t_0|$ will have the 
same values of $\Lb$. However, these objects will have very different SFRs because those with $t>t_0$ will include
the peak of SF within their SFRs, while for $t<t_0$ the SFR will mostly include the secular SF. 
This effect is more severe for larger values of $|t-t_0|$, and once the burst duration is smaller than
the time-scale of SFR, contributing significantly to the scatter at low $\Lb$.

In Fig.~\ref{fig:sfr_var} we demonstrate these effects on our model galaxies at $\z=0$. We first plot 
individual SFRs and AGN luminosities from the model, and show how the average SFR is flat 
at $\Lb<10^{44}$ erg s$^{-1}$. We then plot for the same objects the SFRs that are 
calculated based on bursts only, taking out the contribution from the secular SF mode. 
The SFR due to the bursts is more correlated with the AGN luminosity. 
The remaining scatter between the bursty SFR and $\Lb$ originates from the fact that 
AGNs can be selected both before and after the peak (i.e. $t_0$ can occur before or after the observation). 
This changes the contribution to SFR from the burst itself, although the AGN luminosity is left unchanged. 
In Fig.~\ref{fig:sfr_var} we show in different symbols AGNs that are selected before the peak of
the burst, with $(t-t_0)/\sigma_b<-1$. For these objects, the correlation between SFR and AGN luminosity is 
strong, as expected. The strength of this last effect is smaller when $\sigma_b$ is higher (even when
using $\sigma_0=0.2$ instead of the current value of 0.13, the scatter due to different values of $x$
becomes less dominant).

Our model suggests that 
only $\sigma_0$ and $\fb$ affect the rates diagram. Changing the value of $\sigma_0$ results
in modifying the average level of SFR for low $\Lb$. On the other hand, $\fb$ shifts the diagram
along the X-axis. Unlike the behaviour of the luminosity function, 
there is no simple degeneracy between these parameters here. 
The value of $\sigma_0=0.13$ Gyr that is chosen for the best model is obtained by 
matching the observed rates diagram.

We have tested how other standard SF modes affect the rates diagram
and found that the $\Lb$-SFR  correlation is very sensitive to secular modes. 
For example, when using accretion into BHs that is proportional to the secular SFR, 
there are no regions with uncorrelated SFR and AGN luminosity. This is apparent even in cases where the 
BH growth in the secular mode is only $10^{-5}$ of the assumed SFR.
As indicated by Fig.~\ref{fig:sfr_var}, the rates diagram originates from two effects. First, there is
hardly a correlation between the SF within the burst and the secular SF. Second, the episodic nature
of the burst together with the integration time of the SFR indicator. 
These two effects do not exist in a secular mode that is based on continuous BH accretion mode. However, 
different types of  secular accretion, e.g. episodic events that were not tested by us, might show 
better agreement with the observed rates diagram.

The rates diagram as shown here is defined as the average SFR at a given AGN luminosity, and
is not affected by the number of AGNs of different luminosities. However, once we compute the average
AGN luminosity as a function of SFR (dashed line in Fig.~\ref{fig:sfr_var}), we see a different trend, in which
SFR and $\Lb$ are always correlated. This way of averaging hides the flat
part of the diagram, occupied by low luminosity
AGNs, and is more dependent on the number of low-luminosity AGNs (i.e. on sample selection). 
Since the flat part of the rates diagram is highly restricting the models, the average $\Lb$
 is less restrictive when comparing models to observations.

In Fig.~\ref{fig:sfr_burst_dt} we plot for each value of SFR and AGN luminosity the median value of 
$x\equiv(t-t_0)/\sigma_b$. According to Eq.~\ref{eq:merger_time}, $x$ is the location within the Gaussian of the burst,
where negative values (down to -2) correspond to the burst starting point, and 
positive values (up to 4) correspond to the
late episode of the burst. Each burst starts on the left part of the rates diagram, with small AGN luminosity that is 
due to a value of $x=-2$. As $x$ increases towards 0, both the SFR and $\Lb$ rise. Later on, when $x>0$ 
$\Lb$ goes down to its initial level, while the SFR stays a bit higher, due to the integration over
150 Myr.  As a result, the right part of the diagram is mostly populated with $x<-1$, and the higher values of $\Lb$
are related to $x\sim0$.
This effect modifies the correlation of ${\rm SFR} \propto \Lb$ (that is used by the model, Eq.~\ref{eq:bh_growth}) 
such that ${\rm SFR} \propto \Lb^{0.7}$, as observed by \citet{Netzer09}.

\subsection{Merger fraction}
\label{sec:merger_results}

\begin{figure}
\centerline{ \hbox{ \epsfig{file=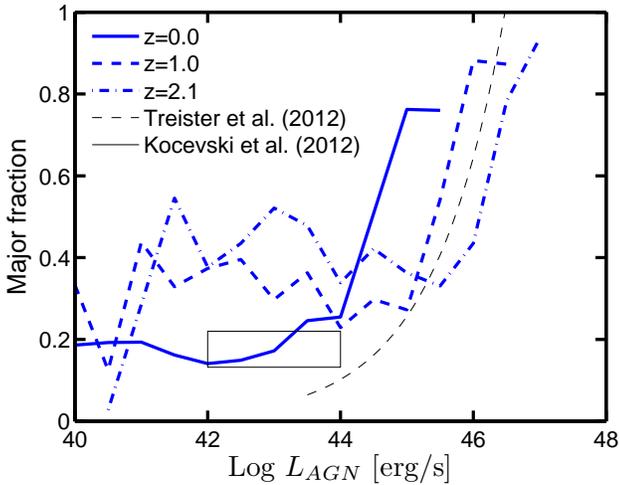,width=9cm} }}
\caption{The fraction of galaxies that host AGNs and are going through major mergers (mass ratio is bigger than 0.3)
out of all galaxies that host AGNs.
\emph{Thick} lines show the model results at various redshifts as indicated, after selecting all galaxies
that host BHs with masses $\log \mb/\msun>7.7$. The \emph{thin} dashed line
shows the fitting function from \citet{Treister12}, obtained from a compilation of various studies at
$0<\z<1$. The \emph{grey} boxy line shows the results from \citet{Kocevski12}, derived at $\z\sim2$. }
\label{fig:mass_ratio}
\end{figure}

\begin{figure}
\centerline{ \hbox{ \epsfig{file=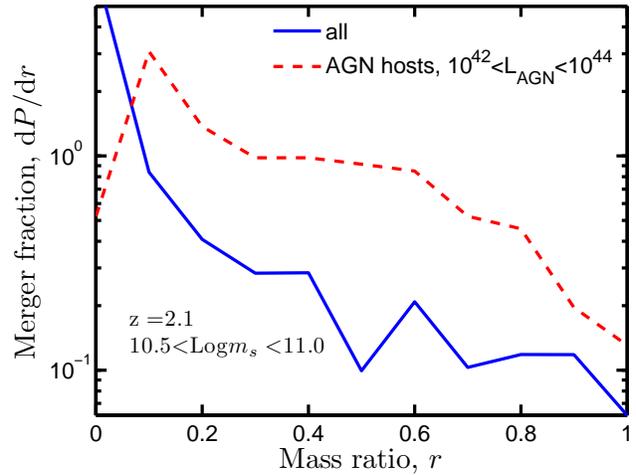,width=9cm} }}
\caption{The fraction of galaxies undergoing a merger event with a mass ratio $r$. d$P$/d$r$ is the probability
distribution of galaxies of different $r$ values. All galaxies are selected
at $\z=2.1$ with stellar mass as indicated, in order to avoid resolution issues. The
\emph{solid} line corresponds to all the galaxies within the model. The \emph{dashed} line shows the fraction
out of all the galaxies that host an AGN with $10^{42}<\Lb<10^{44}$ erg s$^{-1}$, as were selected by
\citet{Kocevski12}.}
\label{fig:morph}
\end{figure}

The main argument against a model of BH growth that is based only on mergers is the morphology of 
AGN hosts. As discussed in the Introduction, numerous works show that most AGNs reside in normal disk
galaxies, with no sign of interaction. Here we test the observational constraint in this respect
more closely. 

In Fig.~\ref{fig:mass_ratio} we show the fraction of galaxies that host AGNs and are participating in 
major merger events, out of all galaxies that host AGNs. Major merger events are defined as those with mass
ratio that is bigger than 0.3. Note that in the SAM, if
a galaxy have more than one merger event, the mass ratio
is determined by the event with the largest contribution to the SFR. A galaxy is considered to be a part of 
a merger event throughout the burst.

Fig.~\ref{fig:mass_ratio} shows that the
results of the model are roughly consistent with those of \citet{Treister12},
although they deviate from the analysis of \citet{Kocevski12}. 
The model predicts that major mergers are associated with more luminous AGNs.
There are a few limitations in this comparison that
should be mentioned. First, we assume that the mass ratio for major mergers is 0.3, which is somewhat 
arbitrary. Second, we assume that the time-scale for detecting the morphological features within the galaxy 
is similar to the assumed duration of the burst. Third, we do not allow for a time-delay between the 
change in morphology and the onset of accretion
into the BH. All these issues complicate the interpretation of the results.

A different way to examine this issue is to compare galaxies that host AGNs to a control sample of 
galaxies with the same stellar mass, that do not host AGNs. According to \citet{Kocevski12}, these two
populations show similar fractions of galaxies that are detected as going through merger
events. In Fig.~\ref{fig:morph} we make the same comparison with our model galaxies. This plot shows 
the fraction of galaxies going through a merger event of mass ratio $r$, after selecting only galaxies
at $\z=2$, and with stellar mass $10.5<\log (m_s/\msun)<11$. The AGN sample includes galaxies hosting
an AGN with $10^{42}<\Lb<10^{44}$ erg s$^{-1}$, similarly to the sample of \citet{Kocevski12}.
This result shows that the fraction of galaxies going through a merger event is 
larger by a factor of $\sim2$ for galaxies that host AGNs. 
Unlike the comparison made in Fig.~\ref{fig:mass_ratio}, here the uncertainty is due to
the existence of a time-delay between changes in morphology and the BH accretion phase.

\subsection{Bulge and black hole mass}
\label{sec:bulge_bh}

\begin{figure}
\centerline{ \hbox{ \epsfig{file=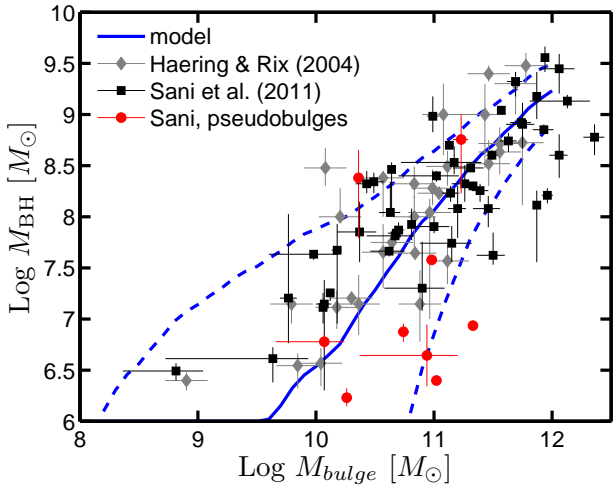,width=9cm} }}
\caption{The relation between the masses of bulges and BHs at $\z=0$. \emph{Symbols} correspond to the
observational results from \citet{Haering04} and \citet{Sani11}. The \emph{thick} solid line
shows the median BH mass for a given bulge mass, taken from the model galaxies. \emph{Dashed} lines
show the 5 and 95 per cent levels of the BH distribution for each bulge mass.}
\label{fig:bulge_bh}
\end{figure}

In Fig.~\ref{fig:bulge_bh} we show the relation between 
bulge and BH mass in our model. The ratio of the two masses depends on various factors. First,
bulges can grow in major merger events by using the stellar
mass from the disks of both progenitor galaxies. This effect will allow bulges to gain more mass than
the total SFR during mergers, more than what is used for feeding the BH. In addition, the 
scatter between the bulge and BH masses should be a result of $\fp$, which allows for only a part of the burst
events to feed the central BH. 

Our model agrees with the results of both \citet{Haering04} and \citet{Sani11} 
at redshift zero. However, the recent study by \citet{McConnell12} shows that at a 
bulge mass of $10^{12}\,\msun$ the BH mass is higher than what we get here, reaching $10^{10}\,\msun$.
Our model BHs are less massive at the high mass end because we use the mass function from \citet{Shankar09}
for calibrating the mass of BHs in our model. In addition, the model predicts a smaller
scatter in the mass relation for more massive objects in comparison to observations. 

\subsection{The distribution of BH masses for a given AGN luminosity}
\label{sec:mbh_L}

\begin{figure}
\centerline{ \hbox{ \epsfig{file=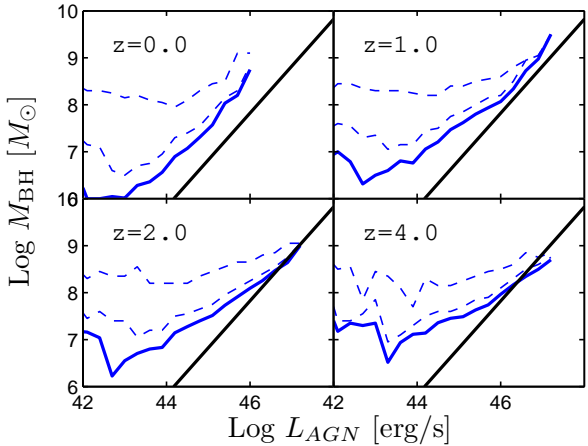,width=9cm} }}
\caption{The distribution of BH masses as a function of AGN luminosities at various redshifts.
The \emph{thick solid} lines are the median values of the model 
The \emph{dashed} lines show the 67 and 95 per cent levels of the distribution.
The thick \emph{straight line} in each panel shows the Eddington luminosity, Eq.~\ref{eq:Ledd}.}
\label{fig:M_L}
\end{figure}

The distribution of BH masses, for a given value of $\Lb$ are shown in Fig.~\ref{fig:M_L}. 
The model predicts that most AGNs accretion is below the Eddington limit, especially at low redshift.
At higher redshifts and high $\Lb$ the AGNs accretion is closer to the Eddington limit.
The scatter we use for the Eddington threshold allow for some objects to slightly exceed the formal
Eddington limit, mostly at high redshift.

\subsection{Specific SFR in AGN hosts}
\label{sec:ssfr}

\begin{figure}
\centerline{ \hbox{ \epsfig{file=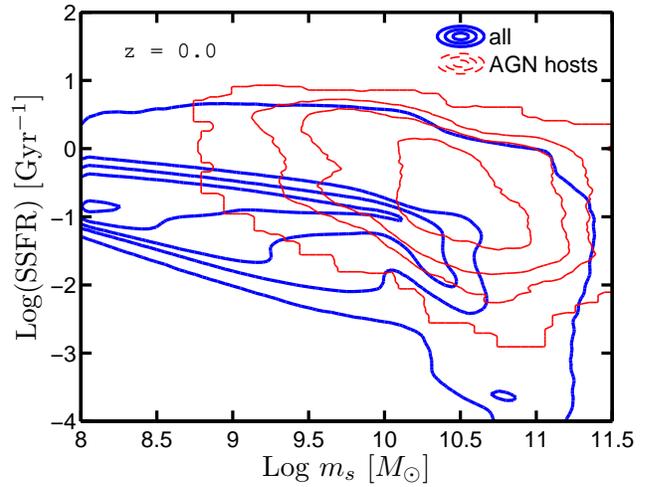,width=9cm} }}
\caption{The distribution of specific SFR (SSFR) versus stellar mass at $\z=0$.
\emph{Thick} contours represent the distribution of all the model galaxies,
while \emph{thin} lines show the distribution of galaxies that host AGNs with $\Lb>10^{42}$ erg s$^{-1}$.
Contour levels are at $\log P=-2.9,\,-1.4,\,-0.9,\,-0.4,\,0.1$ where the integral of $P$ equals
unity.}
\label{fig:ssfr}
\end{figure}

Interestingly, our merger scenario naturally predicts that AGNs are hosted by massive star-forming
galaxies. This can be seen in Fig.~\ref{fig:ssfr}. Observations of this type were done by \citet{Salim07}
and \citet{Bongiorno12} using different SFR indicators. This diagram is an extension of the rate diagram
shown in Fig.~\ref{fig:sfr}, as it shows the full distribution of specific SFR, for each value of stellar mass.
Results at higher redshifts are similar to what is shown here for $\z=0$.

\subsection{BH mass and accretion rate in the SFR-$\Lb$ plane}
\label{sec:sfr-L_plane}

\begin{figure}
\centerline{ \hbox{ \epsfig{file=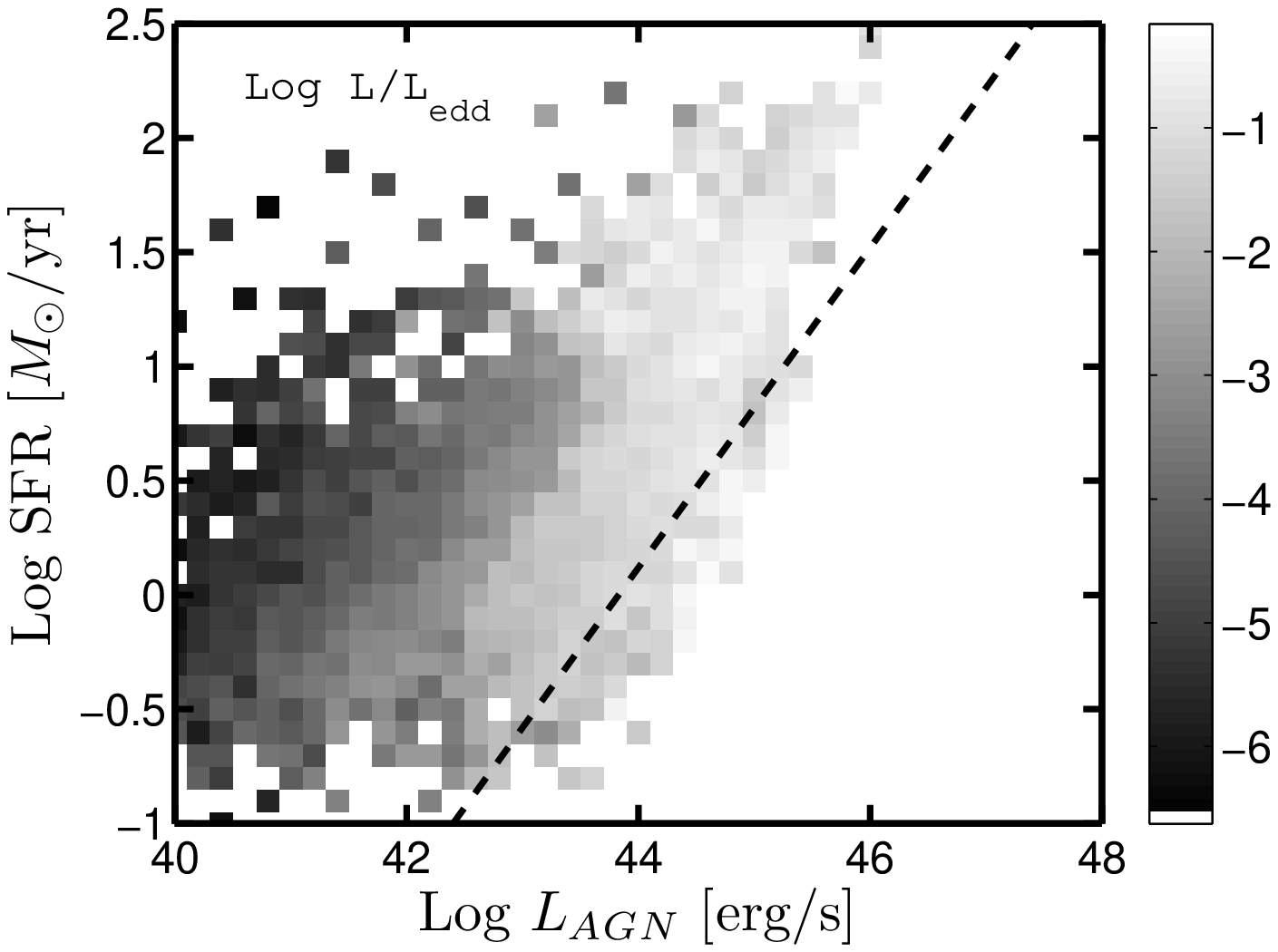,width=9cm} }}
\centerline{ \hbox{ \epsfig{file=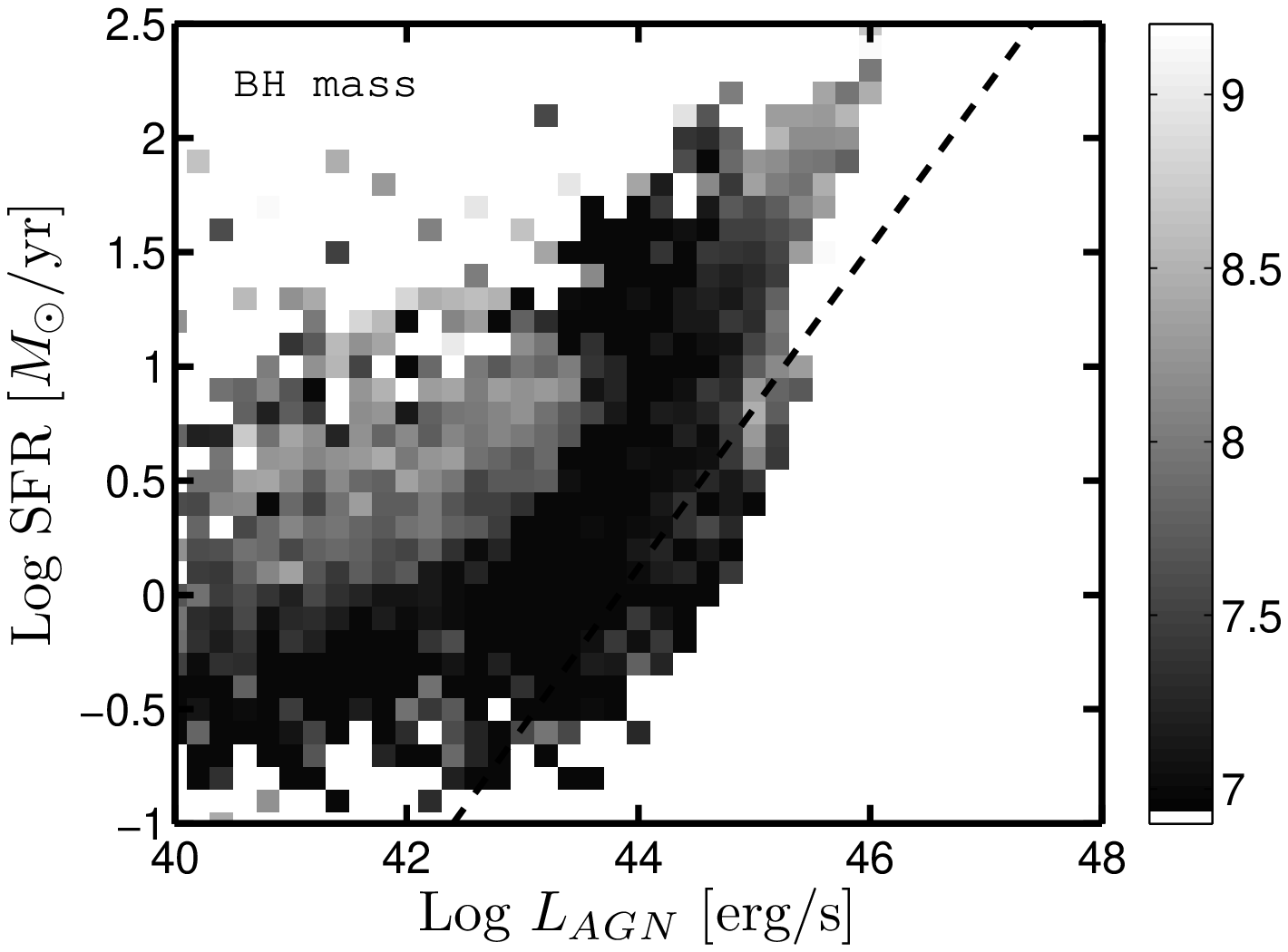,width=9cm} }}
\caption{SFR versus $\Lb$ at $\z=0$.
The \emph{upper panel} shows the median Eddington ratio, $\log \Lb/L_{\rm Edd}$, at each bin. The \emph{lower panel}
represents the median BH mass in units of $\log\msun$. Here we set BH masses to a minimum value
of $10^7\,\msun$.}
\label{fig:sfr_edd}
\end{figure}

In Fig.~\ref{fig:sfr_edd} we show how other properties of AGNs depend on both SFRs and AGN luminosities at $\z=0$.
We show that the model predicts a higher values of $\Lb/L_{\rm Edd}$ at higher $\Lb$, with small dependence on SFRs.
This is in spite of the lower values of $x=(t-t_0)/\sigma_b$ seen in Fig.~\ref{fig:sfr_burst_dt} at high
$\Lb$ and low SFRs. As a result, the Eddington ratio does not depend on $x$, the time within the burst.
We have further tested that the Eddington threshold does not play a role at shaping the rates diagram at
$\z=0$.

In the lower panel of Fig.~\ref{fig:sfr_edd} we show the median BH mass at each value of SFR and $\Lb$.
The scatter in masses around $\Lb<10^{44}$ erg s$^{-1}$ depends on the SFR but other regions of the diagram 
include more uniform distribution of BH masses.

\section{Summary and conclusions}
\label{sec:discuss}

This work presents a semi-analytic model (SAM) for the formation of galaxies and black holes (BHs) within
a $\Lambda$CDM Universe. Our model is built on a galaxy-formation model that matches various observations of galaxies.
These include the stellar mass function at $0<\z<4$, the distribution of star-formation rates (SFRs) at $\z=0$, and
the auto-correlation function of galaxies at $\z=0$. The results were developed in a previous study \citep{Wang12}, 
and are adopted here with no further modifications to the properties of galaxies. When adding BHs to the model,
we are interested in the properties of BHs and active galactic nuclei (AGNs), and how they correlate with the 
properties of their host galaxies. We do no take into account the possible feedback from AGNs on the physics of gas
within galaxies.

Our model assumes that BHs grow only during star-formation bursts, with a probability $\fp$ that is the same for 
all galaxies at all redshifts. The bursts in our model are only due to merger events; we do not model bursts from
secular processes such as disk instability. When an event of BH accretion occurs,
we assume that the gas mass that reaches the BH equals a fraction $\fb$ of the stars made in the burst at the same 
time. We use two additional parameters to describe the threshold on accretion due to the Eddington 
luminosity, in order to limit the accretion in case the burst is large and the BH mass is small. 
This model for the formation and evolution of BHs is therefore very simple, also in comparison to previous
SAMs \citep{Kauffmann00, Croton06, Bower06, Malbon07, Monaco07, Somerville08, Bonoli09, Fanidakis11, Fanidakis12,
Hirschmann12}.

The observational constraints used here are the mass function of BHs at redshift zero \citep{Shankar09}, the
luminosity function of AGNs at $\z\lesssim4$ \citep{Richards06,Croom09}, and the relation between SFRs and AGN 
luminosities \citep{Shao10, Rosario12}. We specifically study the latter constraint, as it was not tested by 
previous models. According to \citet{Rosario12}, AGNs at low redshift are related to a constant level of SFR 
within their host galaxies for $\Lb<10^{44}$ erg s$^{-1}$. At higher AGN luminosity, 
where $\Lb$ is larger than the luminosity of SFR, a power-law type
correlation, ${\rm SFR}\propto\Lb^{0.7}$ is seen \citep[see also][]{Netzer09}. At $\z\sim2$, 
AGNs of all luminosities are hosted within galaxies of the same average SFR, at a level of $\sim30$ $\msun$ yr$^{-1}$.
We term these results as the `rates diagram', since it shows the relation between the star-formation rate, and the
rate of BH growth.

As was
demonstrated in Figs.~\ref{fig:sfr} \& \ref{fig:sfr_var}, our model fits well the observed rates diagram.
A necessary condition for the agreement is that we adopt the correct time scales for both SF and AGN activity. 
We therefore assume that the luminosity at $60\mu m$  used by \citet{Rosario12} represent an 
averaging of 150 Myr over the SFR of each galaxy, and that the AGN luminosity is instantaneous (i.e. using the minimum 
time-step of our model, $\sim10$ Myr). When using these time scales, we naturally 
reproduce the observed rates diagram. In our model, the lack of correlation between SFR and AGN luminosities is due to
two main reasons: First, secular SF in galaxies are not associated with BH activity.
Second, BH accretion before and after the peak in SFR could have the same AGN luminosity,
but different SFR. A key assumption in the model is the lack of correlation between the
properties of a galaxy prior to the merger event, and the amount of star-formation within the burst, which depends
on the properties of both merging galaxies. We thus argue that the rates diagram favours a mode of BH accretion
that is due to mergers.

The rates diagram is used here for constraining the duration of the bursts in our model and for constraining the BH 
accretion factor, $\fb$. The diagram is very sensitive to secular modes of BH growth (i.e. 
modes that do not depend on merger events). Even a small level of BH accretion from the secular mode, which adds a 
negligible mass to BHs, can modify the rates diagram considerably. Although we do not model such scenarios here,
more constraints could be obtained by measuring the SFR in the central regions of galaxies \citep[e.g.][]{Stern12}.

Visual classification of AGN hosts indicates that most AGNs are located within normal disk
galaxies, with no signs of mergers \citep{Gabor09, Cisternas11, Kocevski12, Schawinski12, Treister12}. We show
in Figs.~\ref{fig:mass_ratio} \& \ref{fig:morph} that a similar result is obtained here, using a model with
BH growth coming only from mergers. This is because most BH accretion is associated with minor merger
events, which are difficult to detect observationally. However, our model shows that
a typical galaxy has roughly half the chance to experience a merger event
in comparison to a galaxy of the same stellar mass that hosts an AGN. We do not test in this work the 
possibility of having a time-delay between the merger event and the episode of BH accretion. Such a time-delay
will further decrease the amount of AGN hosts that are showing morphological signs of interactions.

The model parameters used here were tuned to fit the observed mass function of BHs at $\z=0$,
and the luminosity function of AGNs at $\z\lesssim4$. In addition, our final model agrees with the observed 
correlation between the mass of bulges and BHs \citep{Haering04,Sani11,McConnell12}. We show in Fig.~\ref{fig:ssfr}
that our model predicts that AGN hosts are relatively massive, and star-forming galaxies.

Finally, we provide predictions for the mass of BHs, and their Eddington ratios, within the rates diagram,
as a function of both SFR and $\Lb$. These predictions could be used to test the model 
against new observations.


\section*{Acknowledgments}

The Millennium Simulation databases used in this paper and the web application providing online access 
to them were constructed as part of the activities of the German Astrophysical Virtual Observatory.
EN acknowledges funding by the DFG via grant KH-254/2-1.
We thank David Rosario, Eleonora Sani, and Francesco Shankar for sharing their data
in electronic format; David Rosario, Francesco Shankar, Dieter Lutz, and Benny Trakhtenbrot for useful discussions; and
Sadegh Khochfar for reading an earlier draft and for many helpful suggestions.
%

\bibliographystyle{mn2e}
\bibliography{ref_list}

\appendix
\section{Comparison with previous models}

\begin{figure}
\centerline{ \hbox{ \epsfig{file=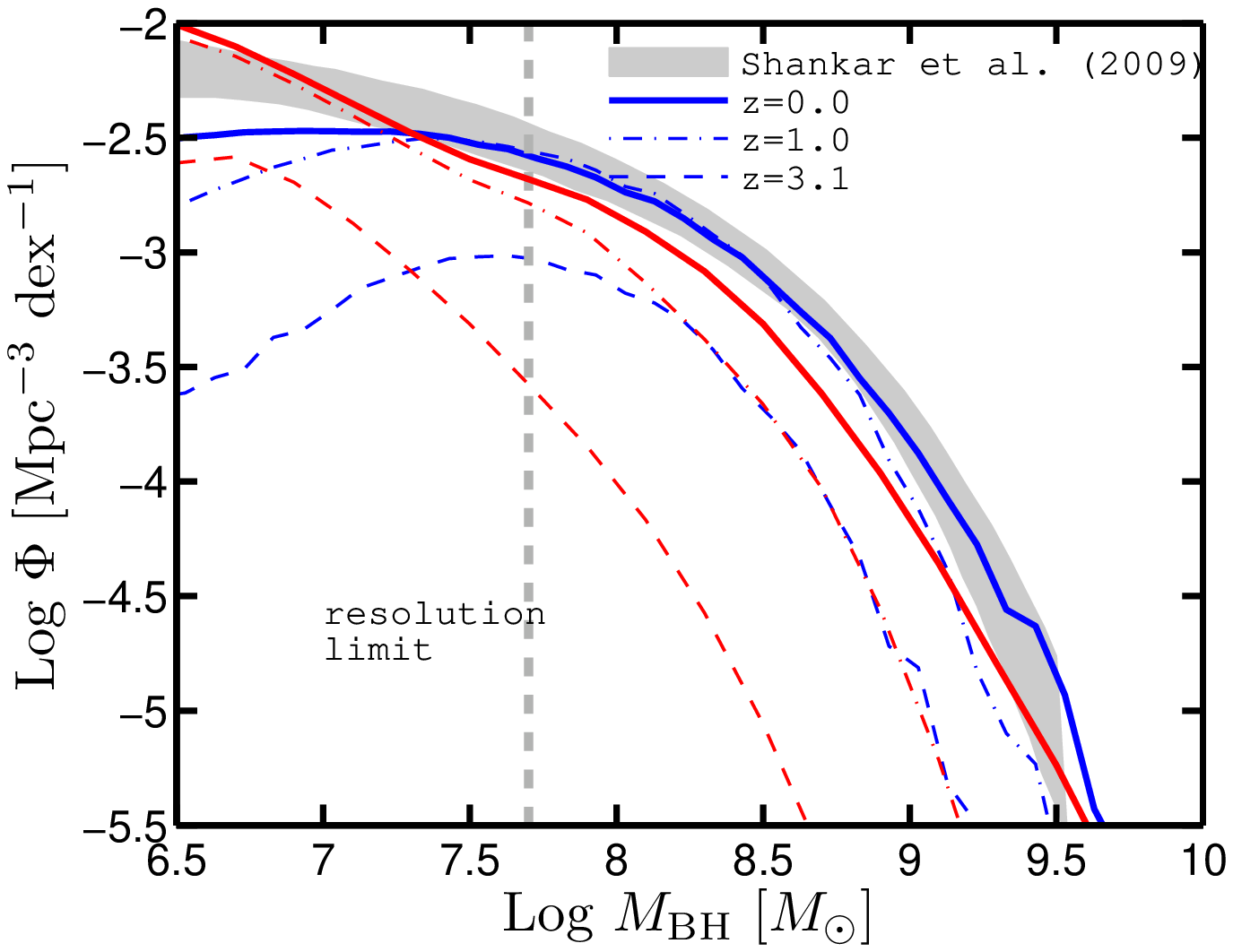,width=9cm} }}
\caption{The mass function of BHs. The \emph{shaded} region represents the observational prediction
from \citet{Shankar09} based on the correlation between bulge and BH masses at low redshift. 
Different \emph{line} types correspond to the model mass function at various redshifts as indicated.
The grey \emph{dashed} line shows the minimum BH mass that is reliably reproduced by the model 
(due to a minimum subhalo mass of $1.72\times10^{10}\,\hmsun$). 
We add in \emph{red} lines results from the public model of \citet{DeLucia07} for comparison. }
\label{fig:bh_mf1}
\end{figure}

As was discussed in the main body of the paper, previous models \citep[e.g.][]{Croton06,DeLucia07,Marulli08}
have obtained different results with models that are often based on the same $N$-body simulation as is used here.
In order to provide more details on the difference between our model and previous ones, we plot in Fig. \ref{fig:bh_mf1}
the mass functions of BHs from our model, along with the results from the publicly avialable model of \citet{DeLucia07}.
It can be seen that our model builds the mass of BH faster at high redshift than the previous model. This is probably
due to the difference in star-formation law that was implemented in our model \citep[see][for more details on our
model]{Wang12}. In brief, our model uses lower SF efficiencies at high redshift, and for low-mass galaxies.
In this way, more cold gas is available for SF during bursts, allowing much larger gas feeding into the central BH.
It is shown in \citet{Wang12} that the cold gas content within our model is much higher than in \citet{DeLucia07}.
It should be mentioned that our model galaxies actually include more cold gas than what is allowed by observations.
This issue, discussed in detail in \citet{Wang12}, should effect the choice of our parameter $\fb$.
Lastly, our merger rates (as a function of galaxy mass) might differ from those of previous models due to
the difference in the stellar mass function. The diffferent BH mass function shown in Fig. \ref{fig:bh_mf1} indicates
that the relation between stellar mass and BH mass is also different in our model, giving rise to different merger
rates \emph{per BH mass} in comparison to previous models.

\label{lastpage}

\end{document}